\documentclass{article}

\usepackage{arxiv}

\usepackage[utf8]{inputenc} 
\usepackage[T1]{fontenc}    
\usepackage{hyperref}       
\usepackage{url}            
\usepackage{booktabs}       
\usepackage{amsfonts}       
\usepackage{nicefrac}       
\usepackage{microtype}      
\usepackage{lipsum}
\usepackage{graphicx, caption, subcaption}
\usepackage[abbreviations,symbols,nogroupskip,nonumberlist]{glossaries-extra}

\usepackage{tabularx}
\usepackage{algorithm,algorithmicx,algpseudocode}
\usepackage{xcolor}
\usepackage{multirow}
\usepackage{comment}

\newcommand{\myparagraph}[1]{\vspace{10pt}\noindent\textbf{#1.}}

\title{Decentralized Collaborative Inertial Tracking}

\author{
 Alpha Diallo \\
  Department of Information Systems\\
  University of Lausanne, Switzerland\\
  \texttt{alpha.diallo@unil.ch} \\
   \And
 Benoit Garbinato \\
  Department of Information Systems\\
  University of Lausanne, Switzerland\\
  \texttt{benoit.garbinato@unil.ch} 
}

\begin{document}
\maketitle

\begin{abstract}

Although people spend most of their time indoors, outdoor tracking systems, such as the Global Positioning System (GPS), are predominantly used for location-based services. 
These systems are accurate outdoors, easy to use, and operate autonomously on each mobile device.
In contrast, Indoor Tracking Systems~(ITS) lack standardization and are often difficult to operate because they require costly infrastructure.
In this paper, we propose an indoor tracking algorithm that uses collected data from inertial sensors embedded in most mobile devices. 
In this setting, mobile devices autonomously estimate their location, hence removing the burden of deploying and maintaining complex and scattered hardware infrastructure.
In addition, these devices collaborate by anonymously exchanging data with other nearby devices, using wireless communication, such as Bluetooth, to correct errors in their location estimates. 
Our collaborative algorithm relies on low-complexity geometry operations and can be deployed on any recent mobile device with commercial-grade sensors. 
We evaluate our solution on real-life data collected by different devices. 
Experimentation with 16 simultaneously moving and collaborating devices shows an average accuracy improvement of 44\% compared to the standalone Pedestrian Dead Reckoning algorithm.

\end{abstract}

\section{Introduction}
\label{introduction}

Positioning systems are increasingly popular due to their ease of use and the growing number of digital consumer electronics such as smartphones, smartwatches, microcontrollers, etc. However, we observe that most of these systems rely on Global Navigation Satellite Systems (GNSS) such as the Global Positioning System (GPS), characterized by their availability, accuracy, and reliability~\cite{o2001availability}. 
Since these systems rely on satellites, they require a direct line of sight with mobile devices. Meaning that they cannot operate in indoor environments. This leaves a gap in potential use cases for location services, knowing that we spend 90\% of our time indoors~\cite{indooratlas}.

\subsection{Indoor Tracking Systems}

Many Indoor Tracking Systems (ITS) have been proposed in the literature and by the industry to overcome the limitations of outdoor positioning systems.
These systems can be classified into two categories depending on whether they require fixed infrastructure.
Infrastructure-based ITSs mostly rely on wireless signals, such as WiFi and Bluetooth Low Energy (BLE), to exchange data or estimate locations. 
Wireless signals are prone to environment perturbation and thus require some aggregation and processing steps, performed mainly by some central server.
In such ITS, the central server acts as a single point of failure, a bottleneck for communications and a potential target for malicious adversaries because it stores sensitive data.
On the other hand, infrastructure-less ITSs mostly rely on inertial sensors embedded in mobile devices and use their processing power to run the tracking algorithms.
Inertial ITSs deployed in infrastructure-less environments suffer from accumulating errors due to environmental perturbations and noisy sensors. 

In the literature, some authors combine inertial and wireless ITS to overcome the above limitations in a so-called sensor-fusion approach~\cite{zou_accurate_2017,el-naggar_indoor_2019}. 
This approach offers better tracking accuracy but does not remove the need to deploy infrastructure, as the devices must calibrate their tracking algorithms using fixed access points.
Another drawback of this approach is the significant computing power required to process a large amount of data several sensors collect.
As a result, this approach is often evaluated in simulated environments or deployed on centralised systems.
Collaborative ITS have been proposed in the literature to overcome the need for infrastructure without losing tracking accuracy.

\subsection{Collaborative Indoor Tracking Systems}

Collaborative Indoor Tracking Systems (CITS) tend to pursue at least one of two goals: (1)~to reduce tracking errors and (2)~to remove the need for a central server.
Using the data collected from nearby devices, these systems effectively track devices while reducing the cost of deploying infrastructures.
In the literature, CITSs can be classified into two categories: centralized and decentralized.

Centralized CITS follow the same architecture as traditional centralized ITS and aim at the first goal, i.e., reducing tracking errors. 
For this, each device is aware of its surroundings and collects data from nearby devices, such as Received Signal Strength Indication~(RSSI). 
The collected data is integrated into the tracking algorithms run by the server to increase the overall accuracy of the tracking solution.
This approach is popular because it gathers tracking data on a central server, allowing for better control. 
Yet it bears the same drawbacks as any centralized solution we already discussed.

In addition to the first goal, decentralized CITSs aim to remove the dependency on a central server by leveraging the computation capabilities of mobile devices to determine their location. 
In this approach, devices exchange data directly with their peers to improve their tracking estimates. 
Decentralized CITS fit well with inertial ITS, as they only need sensors found on most commercial handheld devices, such as smartphones or tablets. 

\subsection{Limitations of existing CITS}

As suggested, most decentralized CITSs in the literature focus on a tradeoff between accuracy and ease of deployment. 
Thus, the problems they address are centred around improving the accuracy of traditional ITS by using collaboration or reducing the hassle of deploying infrastructure and calibrating tracking algorithms.
This is often achieved by relying on wireless technologies, such as BLE beacons with known locations, to track mobile devices~\cite{ho_decentralized_2020}.
However, this approach requires the deployment of numerous BLE~beacons, impacting the deployment and maintenance cost of the solution.

Alternatively, some solutions estimate device locations by exchanging data with nearby devices, using peer-to-peer WiFi beaconing~\cite{noh_infrastructure-free_2018}. 
This approach offers the advantage of not relying on any infrastructure. However, it is impractical with power-constrained devices as WiFi beaconing has a significant impact on power consumption~\cite{putra_comparison_2017}.

In the literature, collaboration is also used by inertial ITS to correct the accumulating errors caused by noisy sensors. 
Qui et~al. propose a collaborative approach with a triangulation calculation model to improve the accuracy of Pedestrian Dead Reckoning (PDR)~\cite{qiu_crisp_2018}. 
Their solution effectively reduces the average deviation from PDR to 0.5 meters.
However, they combine Bluetooth, WiFi and Zigbee technologies, but the latter is absent from most mobile devices. 
In addition, their triangulation calculation model requires collaboration between at least three devices. 
Furthermore, several inertial CITS still require a central server for aggregating data and running the tracking algorithms~\cite{constandache2010did,li_indoor_2013,qiu_crisp_2018}.

\subsection{Contribution and roadmap}

In this paper, we propose a solution that relies on peer-to-peer collaboration to reduce the accumulating errors of inertial tracking.
At the core of our solution lies a typical PDR algorithm, which uses data from inertial sensors embedded in each mobile device to estimate its location. 
This algorithm requires an initial position, heading estimates, and average step length. 
The initial position is typically the last known position the GPS provides before entering a building. 
The heading estimates are collected from the gyroscope embedded in most mobile devices today. 
As for the average step length, it is typically derived from the outdoor displacements of the user. 
The PDR algorithm is then complemented by a collaborative algorithm, which corrects tracking errors thanks to data exchanged anonymously with nearby devices.

Overall, our proposed collaborative algorithm increases the tracking accuracy of typical inertial solutions by 44\%, while at the same time overcoming the limitations of centralized solutions.

The rest of the paper is structured as follows. 
Section~\ref{sec:system_model} describes our system model and introduces more precisely the problem solved by our decentralized collaborative inertial tracking solution. 
Section~\ref{sec:solution} describes our solution in detail.
Section~\ref{sec:experiments} then presents the experimental evaluation.
Section~\ref{sec:discussion} highlights some observations we made after evaluating our proposed collaborative algorithm.
Section~\ref{sec:related_work} discusses solutions comparable to ours.
Finally, Section~\ref{sec:conclusion} summarizes our contributions and sketches possible extensions of this research.
\section{System model and problem}
\label{sec:system_model}

We consider a system composed of an unbounded number of mobile devices, each embedding an Inertial Measurement Unit (IMU) and a wireless communication interface.
The IMU consists of an accelerometer for measuring the linear acceleration of a device in three axes and a gyroscope for measuring the angular velocity.
The wireless communication interface is assumed to be capable of advertising data at regular intervals, with limited impacts on the battery, and detecting nearby devices at a given distance, e.g., by measuring the  received signal strength~(RSS).\footnote{This is typically the case of a BLE interface, which is what we assume for our evaluation experiments in Section~\ref{sec:experiments}.}
Furthermore, each mobile device is assumed to have sufficient memory and computing power to execute the algorithms required by the solution described in Section~\ref{sec:solution}.
Finally, we assume that each mobile has a means to accurately locate itself when outdoors, e.g., via GPS, but not when indoors.

\subsection{Abstract roles}

In order to specify the problem addressed in this paper, we define four abstract roles that are endorsed by the physical components participating in an ITS.
First, we have the \textbf{\emph{trackees}}, which are being tracked.
Second, we have the \textbf{\emph{trackers}}, which are responsible for detecting the presence of the trackees. 
Third, we have the \textbf{\emph{reference points}}, which are responsible for providing reference coordinates within the indoor space.
Finally, we have the \textbf{\emph{aggregators}}, which are responsible for computing the position of the trackees, based on information provided by the trackers and the reference points.

\subsection{Problem specification}

Given the above system model and abstract roles, we can now specify the problem addressed in this paper:
\emph{we aim to build a fully decentralized ITS, where mobile devices are positioning themselves by playing all four roles and by communicating in a peer-to-peer manner to exchange and thus improve their local estimates.} 

More precisely, each mobile device is a \emph{trackee} but also the \emph{tracker} of itself, thanks to its inertial sensors.
Furthermore, each device acts as its own \emph{reference point} when moving from outdoors to indoors and as \emph{reference point} for other devices, when it gets close to them.
Finally, each device acts as \emph{aggregator}, since it computes its own location, using the data obtained through its \emph{tracker} role, and the data received from other devices acting as \emph{reference points}.

\section{A Decentralized inertial CITS\label{sec:solution}}

Our solution is based on two parts. Presented in Section~\ref{subsec:non-collaborative}, the first part consists of a local PDR algorithm that uses the inertial sensors to estimate the position of a device.
The second part, discussed in Section~\ref{subsec:collaborative}, describes the proposed collaborative algorithm that overcomes the limitations of the local PDR algorithm.

\subsection{Local PDR algorithm\label{subsec:non-collaborative}}

The first step for tracking users in inertial-based CITS is to gather data from the devices that are being tracked.
These non-processed data, defined as raw data, are extracted from the inertial sensors. They comprise a timestamp~$t$, the acceleration on three-axis (x, y, z) obtained from the accelerometer at time~$t$, and the angle of gyration obtained from the gyroscope at time~$t$.
To be exploitable, the raw data need to be processed. The processing step starts by computing the magnitude of acceleration using the accelerometer data to detect steps.

The magnitude of the acceleration is defined with the following equation:
\begin{equation}
\label{eq:magnitude}
    mag = \sqrt{x^{2}+y^{2}+z^{2}}
\end{equation}

where $x$, $y$ and $z$ represent the data collected on the three-axis.

To estimate the distance travelled by the user, we use a peak detection algorithm that detects local maxima in the magnitude of acceleration. 
Peak detection algorithms have shown effectiveness in counting steps, thus estimating displacement. 
Brajdic et Harle evaluated this approach for counting steps on unconstrained smartphones and demonstrated an error rate of less than 3\%~\cite{brajdic_walk_2013}. 

After collecting and processing the inertial data, what follows is to run the PDR algorithm. 
The PDR algorithm is expressed with the following equation:
\begin{equation}
    \begin{split}
    X_k = X_{k-1} + S * \cos({\theta})
    \\
    Y_k = Y_{k-1} + S * \sin({\theta})
    \end{split}
    \label{eq:pdr}
\end{equation}

In this equation, $(X_{k}, Y_{k})$ represents the coordinates of the new location. This new location is computed relative to the previous location $(X_{k-1}, Y_{k-1})$, with $S$ representing the step length and $\theta$ being the rotation angle.

\begin{figure}[ht!]
    \vspace{-5pt}
    \begin{subfigure}[t]{0.49\textwidth}
      \includegraphics[width=\textwidth]{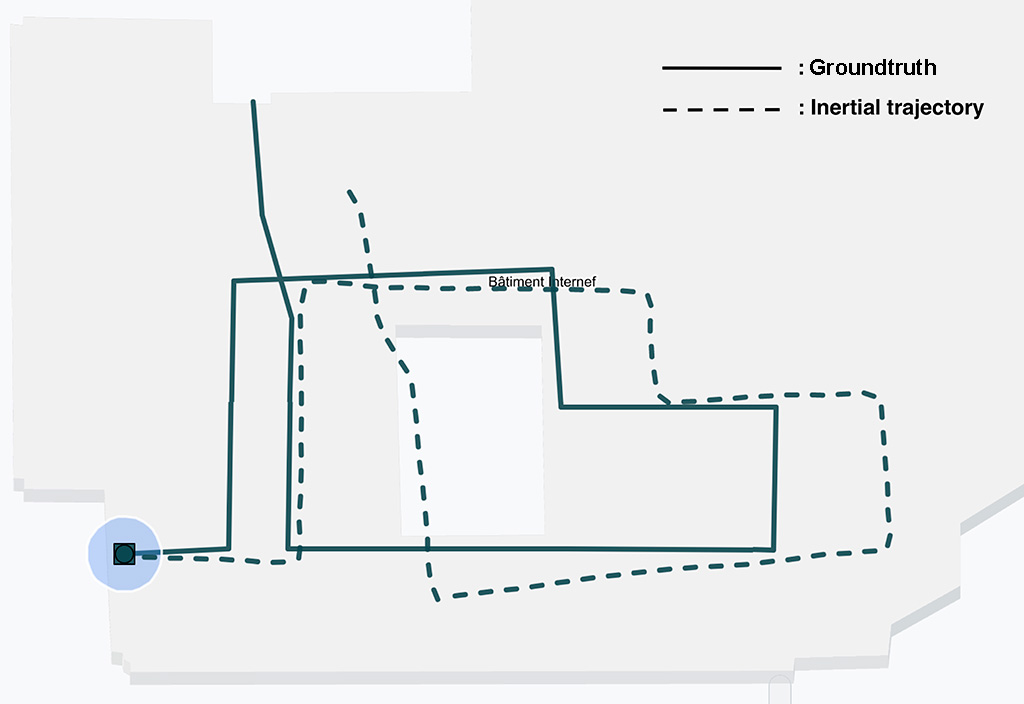}
      \caption{Inertial and groundtruth trajectories}
      \label{fig:drift}
    \end{subfigure}
    \hfill
    \begin{subfigure}[t]{0.49\textwidth}
      \includegraphics[width=\textwidth]{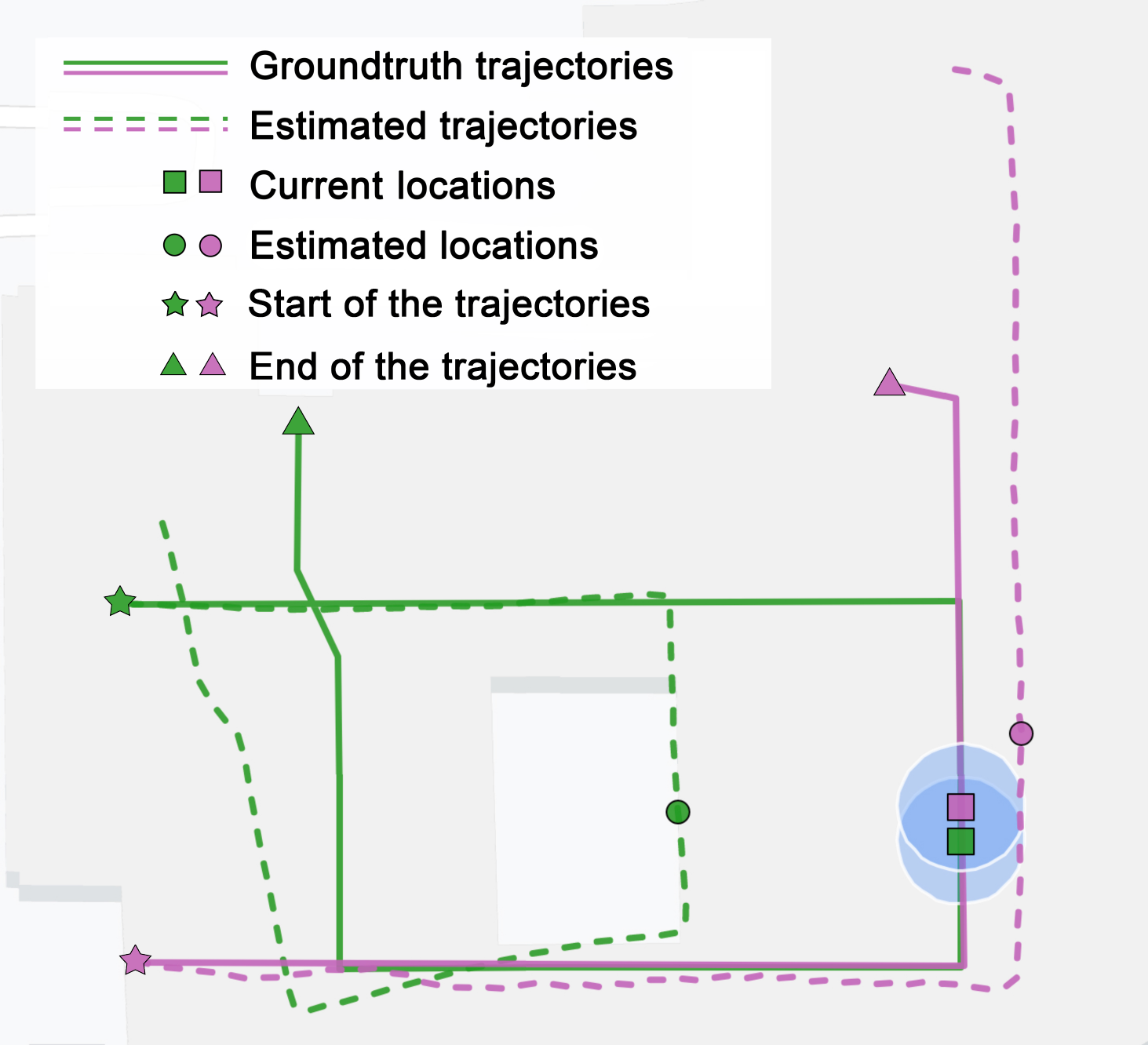}
      \caption{Devices within collaboration distance}
      \label{fig:drift_correction}
    \end{subfigure}
    \caption{Drift generated by the accumulation of errors of inertial sensors.}
    \label{fig:drift+collaboration}
    \vspace{-10pt}
  \end{figure}

The PDR algorithm is known to generate drift because of the accumulation of errors from the inertial sensors. For example, the accelerometer can record noisy data that would be interpreted as a step, thus distorting the calculations. The same applies to the gyroscope prone to environmental perturbations. 
Figure~\ref{fig:drift+collaboration} illustrates this phenomenon. In this figure, the solid lines represent the ground truths, and the dashed lines represent the corresponding trajectories generated with the local PDR algorithm. 
We can observe how the trajectories generated by PDR diverge over time from their respective groundtruths.
In the following, we explain how we reduce the accumulated errors using a collaborative algorithm.

\subsection{Collaborative part of our solution}
\label{subsec:collaborative}

The collaborative approach uses a path-loss model to estimate the distance between nearby devices.
This distance is then used as a filter to consider devices only within close proximity.

A path-loss model estimates these distances using the signal strength between two devices, one acting as an emitter and the second as a receiver.
Theoretically, the signal strength between an emitter and a receiver is correlated with the distance: the shorter the distance, the strongest the signal (and vice-versa). 
The log-distance path-loss establishes this relation with the following equation:
\begin{equation}
  \label{eq:pathloss}
      PL = P_0 + 10\lambda\log(d)+n
  \end{equation}
where $P_0$ is the signal strength at a reference distance, $\lambda$ represents a path loss parameter related to the environment, $d$ is the distance between the trackee and the tracker, and $n$ is a zero-mean white Gaussian random variable.

We estimate the distance by transposing Equation~\ref{eq:pathloss} such as:
\begin{equation}
\label{eq:distance}
    d = 10^{\frac{P_0(d_0)-PL}{10n}}
\end{equation}

\begin{figure}[ht!]
\vspace{-5pt}
\centering
  \includegraphics[width=0.57\textwidth]{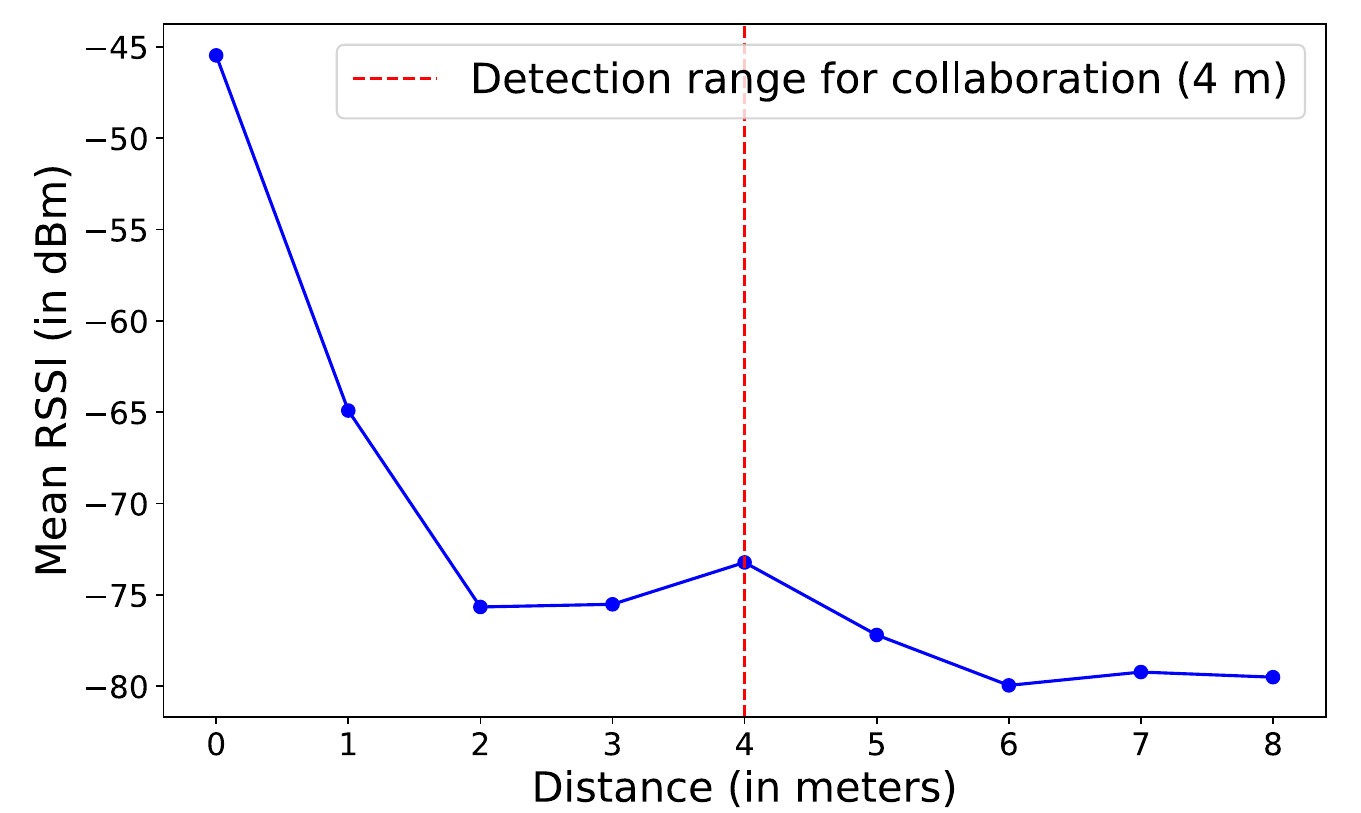}
  \caption{Mean RSSI by distance (in meters)}
  \label{fig:rssi}
\vspace{-10pt}
\end{figure}

For our configuration, we define close proximity as a distance below 4 meters.
This results from an observation that RSSI signals decrease substantially after a certain distance and thus become unreliable. To verify this, we ran an experiment in which we collected measurements from a BLE beacon at a given distance. Figure~\ref{fig:rssi} illustrates our finding and justifies the reason for selecting an RSSI reflecting a distance below 4 meters. 
Indeed, signals collected at a distance above 4 meters fluctuate considerably and, thus, become less reliable.

In addition, as collaboration must occur seamlessly between people with no acquaintance, the distance of 4 meters respects commonly applied social norms defined by the \emph{theory of proxemics}. This theory, developed by Hall et al. in 1966, studies the human use of space within the context of culture. 
Hall et al. define a public space beyond which people will perceive interactions as impersonal and relatively anonymous~\cite{hall1968proxemics}.
According to this theory, a public space around an individual ranges between 3.6 to 7.6 meters.
In Figure~\ref{fig:drift+collaboration}, the distance of 4~meters is represented as blue circles around each device. Figure~\ref{fig:drift_correction} illustrates two devices with overlapping circles, i.e., located within collaboration distance.

As devices move indoors, they broadcast their estimated position and the accumulated errors.
The estimated position is a tuple of two elements composed of their coordinates in a geographic coordinate system. The accumulated errors represent a value that grows linearly over time. 
To process the exchanged data, we introduce a novel collaborative algorithm that takes as input the values described hereafter.
\begin{itemize}
    \item \textit{Devices A and B} are the two devices whose detection ranges overlap. Each device constantly maintains its estimated location and a number representing its accumulated errors.
    
    \item The \textit{lower-threshold} is the level of accumulated errors above which the device starts correcting its estimated location. That is, the collaborative algorithm is executed only when the device errors exceed this predefined threshold.
    \item The \textit{upper-threshold} is the maximum error that is tolerated. When a device has an accumulated error beyond this threshold, it might negatively influence other devices by worsening their estimates. For this reason, its estimated location is no longer used by the devices it encounters until the error starts decreasing at the end of the walking path.
\end{itemize}

When detection ranges overlap, each of the devices run their collaborative algorithm by taking as input data provided by the other device. Thereby, the device running the algorithm is defined as device $A$ and the other device is defined as device $B$. After running the collaborative algorithm, each of the devices updates its own location.

Our proposed algorithm called \emph{Accumulation of Errors (AOE)}, uses the devices' errors accumulated over time to determine how much of the estimations need to be improved. 
It uses a low-complexity geographic computation technique to draw a line between two locations estimated by overlapping devices' detection ranges.
To assign new locations for both devices, the algorithm computes the ratio of accumulated errors for both devices as described in Algorithm~\ref{alg:aoe}. The ratio determines how far the new location would be assigned on the line. In Algorithm~\ref{alg:aoe}, the $intermediatePoint()$ function returns a point whose distance from the starting point on a line is determined by a fraction of the line's length.

With this algorithm, we use stationary devices as reset points by decreasing their accumulated errors whenever they encounter other devices (lines 10 - 12). With this setting, such devices end up having a very low error over time and, thus, significantly impacting other devices.

\begin{algorithm}[ht]
\caption{Accumulation of errors (AOE)}
\label{alg:aoe}
\begin{algorithmic}[1]
    \State \textbf{Input:} devices: $A$ and $B$, lower-threshold $l$, upper-threshold $u$
    \State \textbf{Output:} $A$
    \State $sumErrors \gets A.errors + B.errors$
    \If{$sumErrors \neq 0$}
            \State $ratio \gets A.errors / sumErrors$
            \State $intermediatePoint \gets intermediatePoint(A.location, B.location, ratio)$
            \If{$l < A.errors$ and $B.errors < u$}
                \State $A.location \gets intermediatePoint$
            \EndIf
            \If{$A.location_{t-1} = A.location_{t}$ and $A.errors > 0$}
                \State $A.errors \gets A.errors-1$
            \EndIf
            
        \EndIf
    \State \textbf{return $A$}
\end{algorithmic}
\end{algorithm}

\myparagraph{Data exchange protocol}
A challenge in CITS is how to exchange data collaboratively on smartphones and tablets. A practical and viable solution consists of using BLE, as it offers a mechanism for bidirectional communications, has a low battery consumption, and is widely available. Another advantage of BLE is its low latency while allowing multiple devices to communicate simultaneously. Wu~et~al. experimented with using BLE for exchanging GPS data in a bidirectional manner. They demonstrate that their solution can support at least five devices to communicate simultaneously with each other. In doing so, they measure a  communication latency of 56 ms~\cite{wu_ble-horn_2017}.

In our proposed collaborative algorithm, data that is being exchanged is the estimated locations of the devices $p = (\phi, \lambda)$ and their accumulated errors $e$. Both $\phi$ and $\lambda$ are defined as 64 bits floating-point numbers, while $e$ is an integer that could be stored on 32 bits of memory.
Without any encoding, we can fit the 160 bits (or 20 bytes) in the 255-byte payload allowed by the extended advertising functionality of BLE 5~\cite{badihi_performance_2020}.
\section{Experiments and evaluation}
\label{sec:experiments} 

In this section, we start by describing the experiments, the deployment environment and the devices used for the evaluation. We then evaluate the proposed collaborative algorithm by comparing its results with the local PDR algorithm. 

\subsection{Experimental setting and procedure}

The experiments were conducted in three phases: data collection, data aggregation, and execution replay.

\myparagraph{Data collection}
For the first phase, five participants collected data on a single floor of a university building covering an area of up to $8 400 m^{2}$.
The collection process was done with an Android app deployed on Samsung Galaxy Tabs S7,  which offers an interface to draw the expected walking paths and collect inertial data. It was deployed on tablets to maximize usability, as tablets have larger screens than smartphones. 
The specifications of the devices selected for this experiment match the requirements listed in Section~\ref{sec:system_model}, that is, it embeds an accelerometer, a gyroscope, and a Bluetooth 5.2 chipset.

The data collection process took place according to the following protocol.
Each participant starts by defining their expected walking path with the Android app. 
This is achieved by placing points along the expected walking trajectories that would be followed by the participant. 
This walking trajectory, defined as the \emph{groundtruth}, is used later to compare how each tracking algorithm performs.

During the data collection process, each participant walks at his normal pace while holding their device. 
The Android app collects inertial data from the accelerometer and the gyroscope at a fixed sampling rate of 300 ms (or 3.33 Hz) when the participant walks on the predefined groundtruth. 

\begin{figure}[t]
    \centering
    \includegraphics[height=0.5\textwidth, width=0.5\textwidth, angle=0]{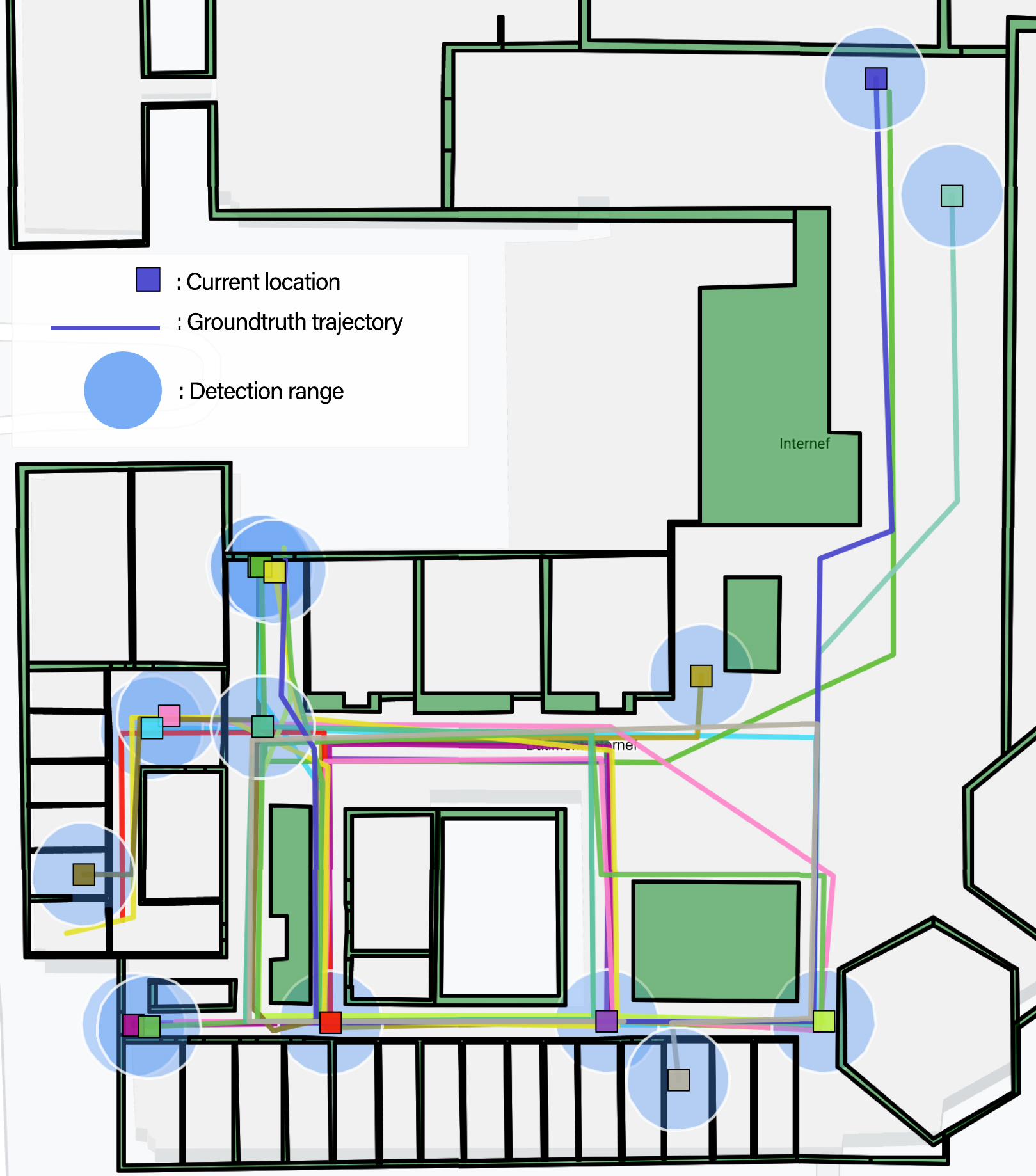}
    \caption{Groundtruth trajectories}
    \label{fig:groundtruths}
    \vspace{-15pt}
\end{figure}

At the end of this phase, we collected 16 trajectories associated with devices and depicted in Figure~\ref{fig:groundtruths}. Each square represents the real location of a participant at a given time. The blue circles represent the detection range of each device, and the solid lines represent the groundtruths.

\myparagraph{Data aggregation}
For the second phase of the experiment, we aggregated the collected trajectories to replay them on the same execution.
As described in Section~\ref{sec:solution}, we started by running the local PDR algorithm to generate the inertial trajectories. 
The parameters required to run the local PDR algorithm are the initial location, the step length and the orientation.

As participants start collecting data from a predefined location, we consider them aware of their initial location and capable of communicating it using the Android app.
Proposing novel approaches for determining an initial location is out of scope as this issue has been studied extensively in the literature~\cite{alzantot_crowdinside_2012,dinh_smartphone-based_2020}. 
For the step length, each participant provides it if known. Otherwise, we estimate the step length by allowing participants to walk on a predefined line with known coordinates.
Then, the line length is divided by the number of steps to obtain an average step length.

In our experiment, participants know their initial heading as they walk through predefined locations. For other configurations, an approach extensively discussed in the literature consists of estimating the initial heading using magnetic sensors embedded in most recent consumer electronics~\cite{li_improved_2017,li_pedestrian_2021}. 

Trajectories computed using the local PDR algorithm are illustrated in 
Figure~\ref{fig:drift+collaboration}. This figure shows how the trajectory generated by the local PDR algorithm accumulates errors over time and rapidly diverges from the groundtruth.

\myparagraph{Execution replay}
For the third phase of the experiment, the testbed is configured to run the proposed collaborative algorithm.
The experiment is done by replaying the real walking paths of each participant, allowing them to move simultaneously, i.e. each participant starts executing their walking path a few seconds after a previous path execution begins. Using the local PDR algorithm, we generate estimated trajectories, and with their detection ranges, each device held by participants detects other nearby devices along their walking path. 
We simulate the BLE data exchange by running the proposed collaborative algorithm when the detection ranges of two devices overlap. 

The execution replay approach allows us to reproduce participants' movements with real-life data and fine-tune the collaborative algorithm parameters. The impacts of these parameters on the proposed collaborative algorithm are described hereafter.

\subsection{Evaluation metrics}
The next step after running the experiments is to evaluate how the proposed collaborative algorithm compares with the local PDR algorithm.
The evaluation is done with two metrics described below.

\myparagraph{Trajectories similarities}
To evaluate how the estimated trajectories diverge from their respective groundtruths, we compute their similarities using the Discrete Fréchet Distance (DFD). It is a measure of similarity proposed by Eiter et Heikki~\cite{eiter1994computing} and described in Equation~\ref{eq:dfd}.

\small
\begin{equation}
	\label{eq:dfd}
	dfd(i, j) = 
	\begin{cases}
		d(P_i, Q_j) & \text{if $i = j = 1$} \\
		max 
		\begin{cases}
			d(P_i, Q_j)  \\
			min 
			\begin{cases}
				dfd(i-1,j) \\
				dfd(i,j-1) \\
				dfd(i-1,j-1)
			\end{cases}
		\end{cases} & \text{otherwise}
	\end{cases}
\end{equation}
\normalsize

where $P$ and $Q$ represent trajectories such as $P = \langle p_1, ..., p_m \rangle$ and $Q = \langle q_1, ..., q_n \rangle$, with $p_i$ and $q_i$ representing points on each of the trajectories. Then, $d(P_i, Q_j)$ is the ground distance $d$ between points pertaining to their respective trajectories $P$ and $Q$. The ground distance between two points $p_{l} = (\phi_{l}, \lambda_{l})$ and $q_{k} = (\phi_{k}, \lambda_{k})$ is computed with the harvesine formula, first proposed by Jamez Andrew in 1805~\cite{dauni_implementation_2019} and defined in Equation~\ref{eq:haversine}.

\small
\begin{equation}
	\label{eq:haversine}
	d = 2 R \arcsin{\sqrt{\sin^{2}\left({\frac{\varphi_{l}-\varphi_{k}}{2}}\right)+\cos(\varphi_{k})\cos(\varphi_{l})\sin^{2}\left({\frac{\lambda_{l}-\lambda_{k}}{2}}\right)}}
\end{equation}
\normalsize
where $R$ is a constant representing the radius of Earth.

\myparagraph{Localization errors}
This second metric represents a pairwise ground distance between individual points of two trajectories.
To compute the accuracy of the local and collaborative algorithms, we use the third quantile of the localization errors. This metric is less prone to outliers than the mean localization errors, and it is now commonly used in the literature as a performance metric~\cite{potorti2017comparing}.

To evaluate a tracking algorithm, the third quantile of errors and the DFD are complementary, as the first offers the advantage of being robust to outliers and easy to interpret. Conversely, the latter is robust to temporal variability when dealing with complex data and considers the entire trajectories by capturing non-linear relationships between points of each trajectory.

\subsection{Evaluation results}

The evaluation shows how our AOE algorithm performs compared to the local PDR algorithm regarding similarity and localization errors. 

\begin{figure}
\centering
  \begin{subfigure}{0.49\textwidth}
    \includegraphics[height=.8\textwidth]{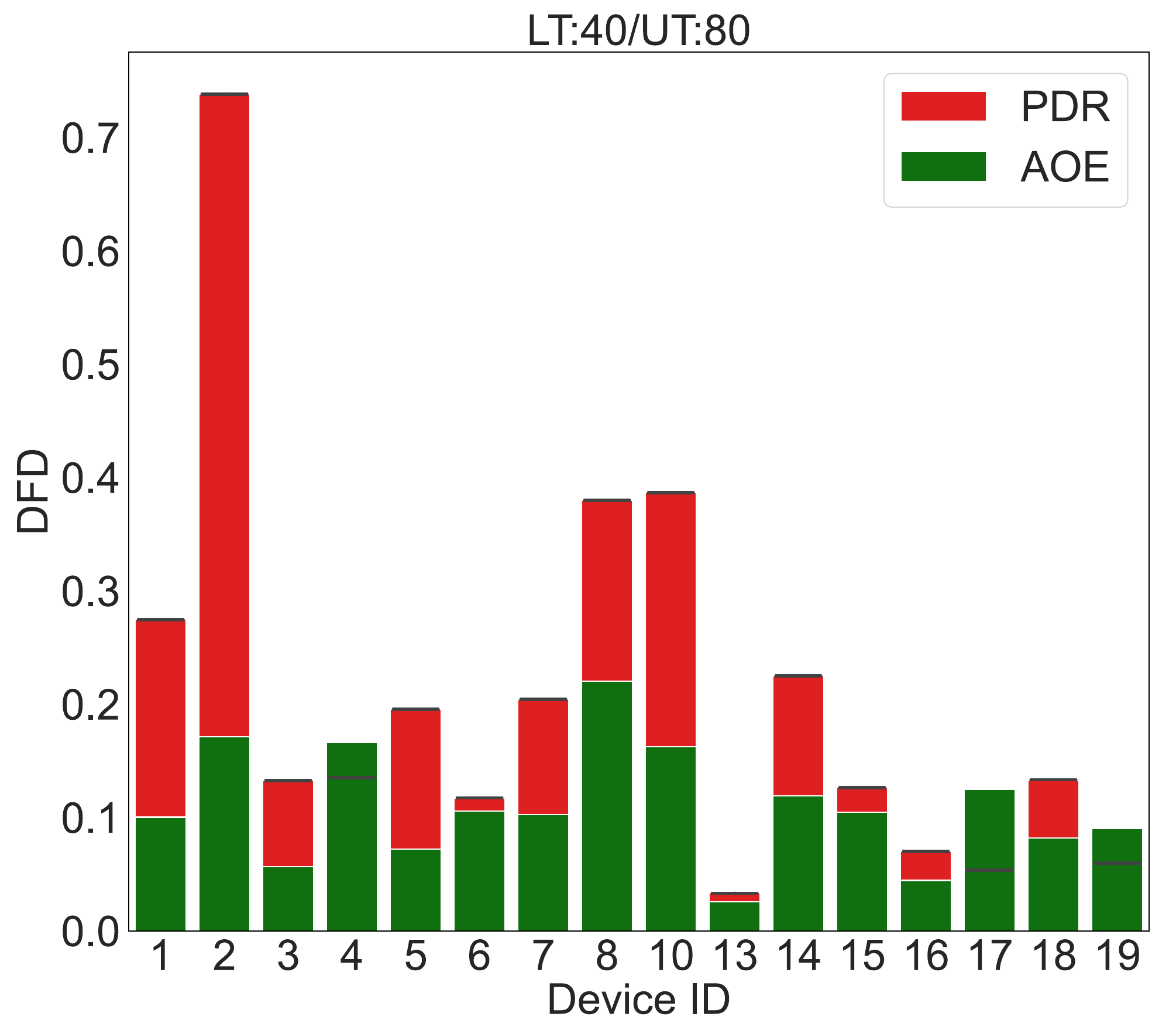}
    \caption{DFD score}
    \label{fig:complete_dfd}
  \end{subfigure}
 \hfill
  \begin{subfigure}{0.49\textwidth}
    \includegraphics[height=.8\textwidth]{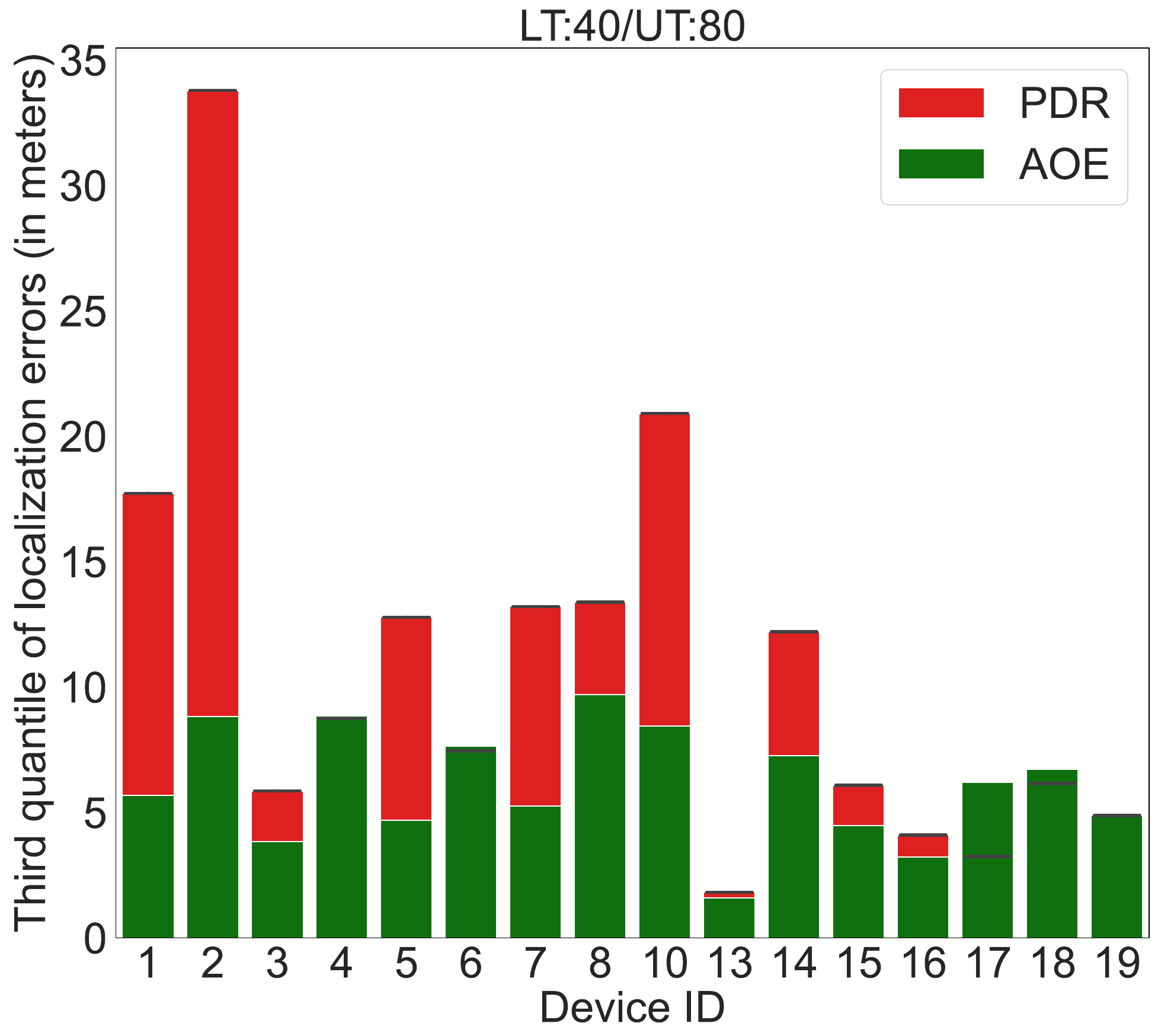}
    \caption{Third quantile of localization errors}
    \label{fig:complete_quantile}
  \end{subfigure}
  \vspace{-4pt}
    \caption{DFD and AOE Algorithms evaluation}
    \label{fig:image2}
    \vspace{-15pt}
\end{figure}

\myparagraph{Trajectories similarities}
For collaborative algorithms, it is essential to highlight the similarity as some devices might negatively impact their peers, usually in the early stage of the collaboration, before progressively correcting their estimates.
For our experiments, as observed in Figure~\ref{fig:complete_dfd}, the AOE algorithm substantially increases the similarities of 13 devices out of the 16.
In this plot, the x-axis represents devices identified by their unique identifier. The y-axis represents the DFD of each trajectory generated by a device. The lower the DFD, the higher the similarity with the groundtruth.
The red bars are the DFD of the local PDR algorithm, and the green bars are the DFD of the AOE algorithm.

\begin{figure}[t]
    \centering
    \includegraphics[scale=0.265]{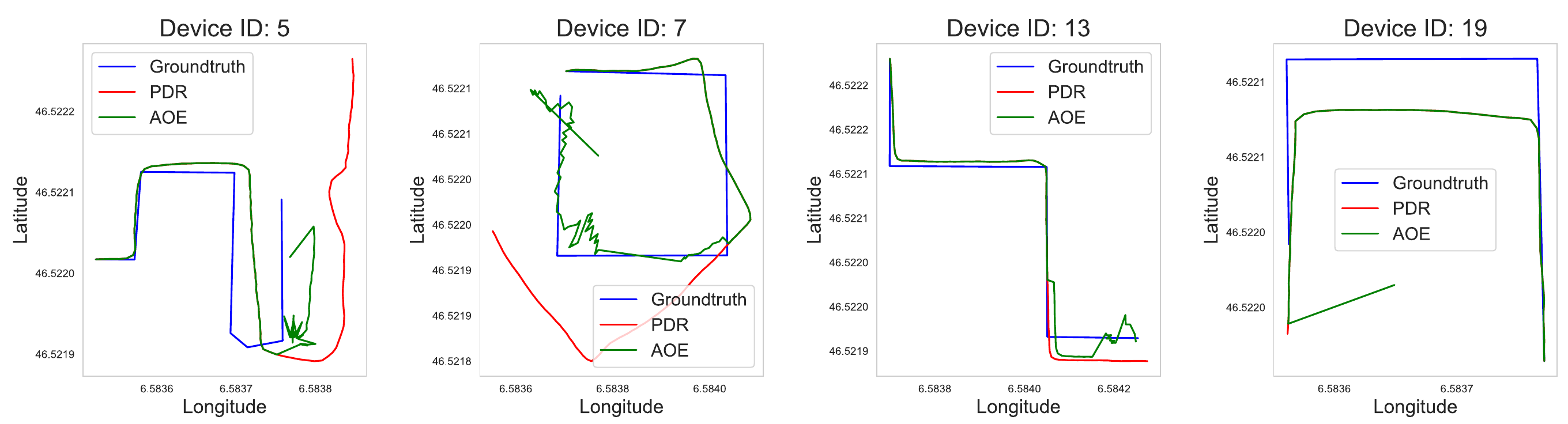}
    \caption{Sample groundtruths and their respective trajectories generated by PDR and AOE algorithms}
    \label{fig:corrected_trajectories}
\end{figure}

Figure~\ref{fig:corrected_trajectories} shows a sample of trajectories generated with the local PDR algorithm (in red), the AOE algorithm (in green) and their respective groundtruths (in blue).
We observe how successive collaborations bring the estimated locations closer to their groundtruths.

\myparagraph{Localization errors}
During the collaboration, some devices tend to deviate from their actual locations after encountering other devices with wrong estimates. This behaviour tends to decrease the localization accuracy between actual and observed locations at a given time. 
However, as mentioned previously, the third quantile of localization errors does not consider the marginal points that deviate significantly from their actual trajectories.

Figure~\ref{fig:complete_quantile} shows the third quantile of localization errors for all the trajectories. As we observed, the AOE algorithm shows better results than the local PDR algorithm on 11 trajectories out of 16. 
With the remaining trajectories, there is either no improvement or a slightly negative impact of less than 3 meters.

To mitigate the impact of wrong estimates on other devices during collaboration, we used a lower threshold of 40 and an upper threshold of 80. This means that devices start collaborating only after accumulating an error of 40, and they do not consider the estimates of devices with errors above 80. 
In the next section, we explain how we selected these thresholds and how they impact the performance of our proposed collaborative algorithm.
\section{Discussion and future work}
\label{sec:discussion}

In this section, we discuss the role of the thresholds, highlight some observations made after the evaluation, and conclude by showing some of the points not addressed in this paper.

\subsection{Importance of thresholds}

The two thresholds introduced in Section \ref{subsec:collaborative}  play an essential role in collaborative algorithms as they define conditions for collaboration.
In inertial-based ITS, as errors accumulate over time, there is little benefit for devices to collaborate at the beginning of their walking paths. 
On the other hand, devices with a large amount of errors might negatively impact the estimates of other devices. 

\begin{figure}[t]
    \centering
    \includegraphics[height=0.6\textwidth]{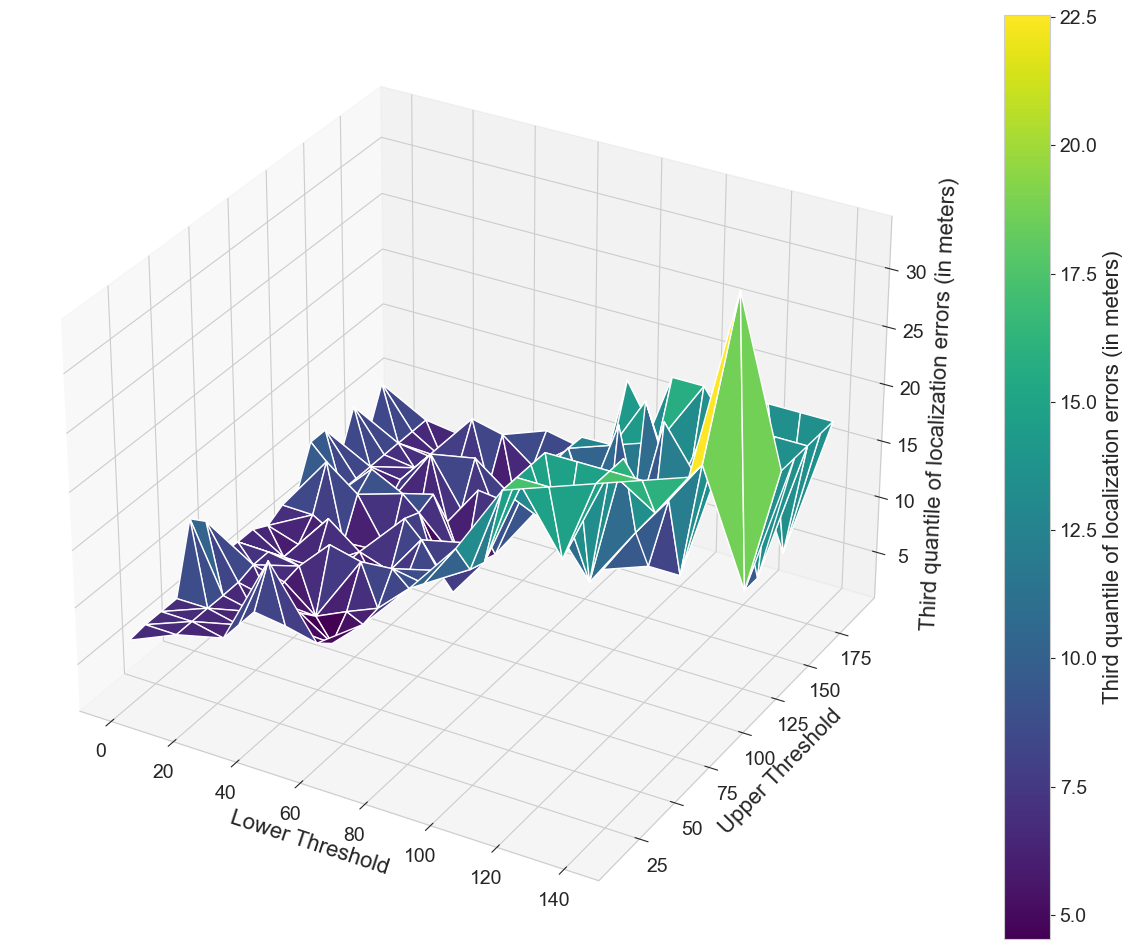}
    \caption{Impact of thresholds on the third quantile localization errors}
    \label{fig:threshold_selection}
\end{figure}

Figure~\ref{fig:threshold_selection} shows the impact of the lower and upper thresholds on the localization errors. As we observe, a high value for the lower threshold inhibits collaboration as few devices are capable of accumulating large errors. In this case, the third quantile of localisation errors of the AOE algorithm is the same as with the local PDR algorithm.

Alternatively, the AOE algorithm shows better results with a small lower threshold, below 60. This means that, for our experiment, devices correct their positions only when they accumulate a certain amount of errors.
We also observe that a low margin between the lower and upper thresholds has a negative impact on the performance of the AOE algorithm.
Since thresholds are preconditions for collaboration, the more restricted these conditions are, the less likely it is for devices to collaborate.

\begin{figure}[t]
    \centering
    \includegraphics[width=.95\textwidth]{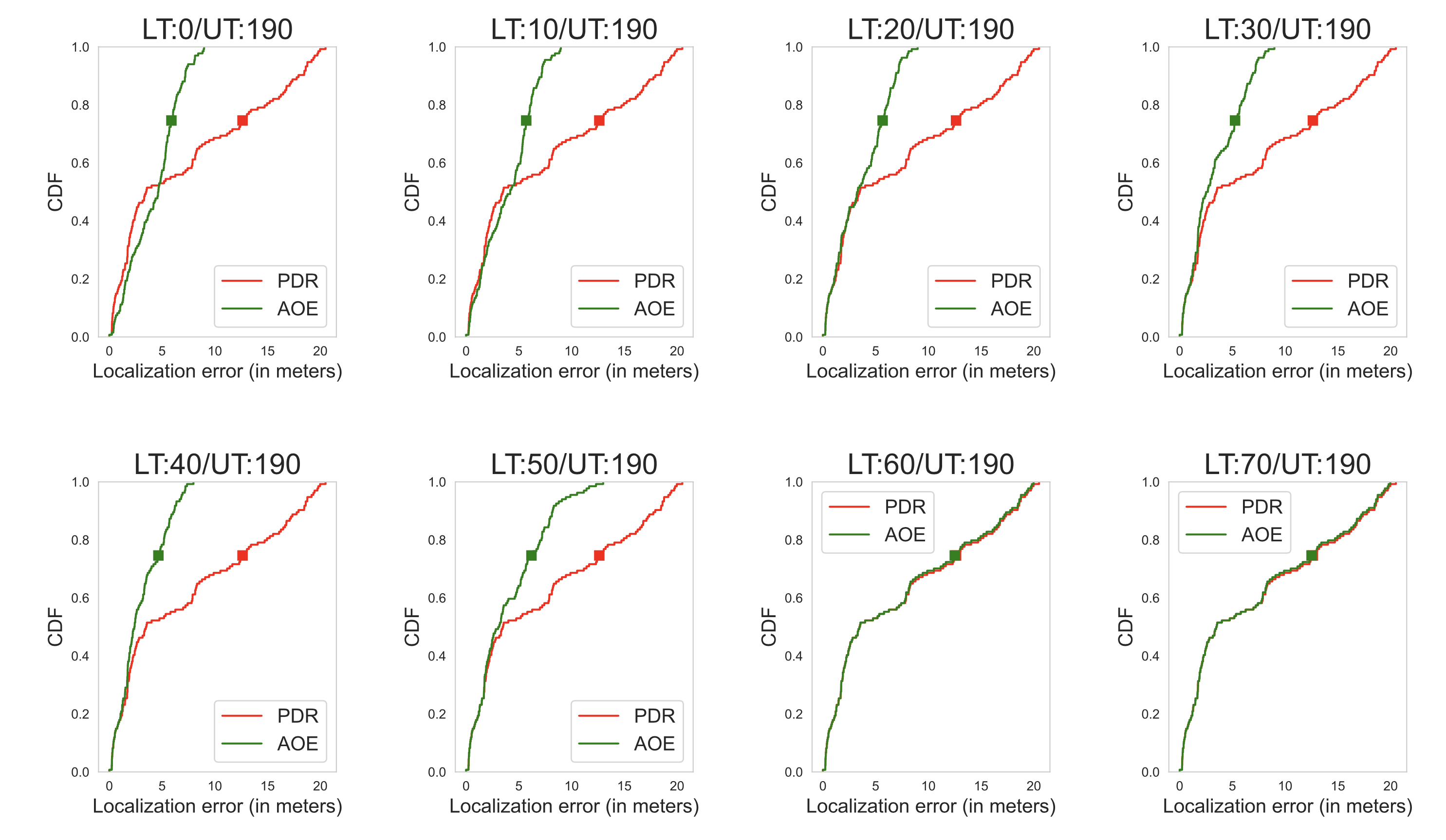}
    \caption{Cumulative Distribution Function of the Localization Errors in meters from a single device}
    \label{fig:localization_errors}
\end{figure}

This can be observed in Figure~\ref{fig:localization_errors}, which presents a Cumulative Distribution Function (CDF) of localization errors (in meters). This figure shows the localization errors of trajectories generated by the local PDR algorithm (in red), and those generated by the AOE algorithm (in green). The squared dots on the plots represent the third quantile of localization errors.

From these plots, we observe that the AOE algorithm significantly improves the trajectories estimated by the local PDR algorithm. This can be explained by continuous collaborations done with a small lower threshold. These plots also show that a high lower threshold reduces the chances of collaboration, thus degrading the accuracy over time. 
At a certain point, when there is no collaboration, the localization errors of the AOE algorithm match those of the local PDR algorithm. Therefore, the estimated trajectories follow the same pattern as the groundtruth.

\subsection{Performance in dense environments}
\label{subsec:performance_in_dense_environment}

During our experiments, we observed that the AOE algorithm performs well in dense environments with higher chances for collaborating. 
Figure~\ref{fig:corrected_trajectories}  shows four sets of trajectories obtained after running the local PDR and the AOE algorithms with a lower threshold of 40 and an upper threshold of 80.
During the execution of their walking path, Device 5 made 140 collaborations with other devices. Device 7 made 123 collaborations, whereas Device 13 made 53 collaborations, and only 29 were made by Device 19.
As observed with Devices~5 and~7, the higher the number of collaborations, the higher the similarity with the groundtruth. 
For Device 13, most collaborations happened at the end of the execution of their walking path. By decreasing the accumulated errors after a collaboration, Device 13 acts more like a reset point, as its errors tend towards 0. In this scenario, stationary devices positively impact other devices after some collaborations.

An interesting observation comes from the estimates of Device 19, which encounters few devices and starts collaborating at the end of its walking path. These first collaborations negatively impact its estimates by a slight margin. Indeed, the estimates of Device 19 using the AOE algorithm show a DFD score of 0.09, whereas the estimates of the PDR algorithm have a DFD score of 0.05. However, this small difference is not noticeable using the third quantile of localization errors as this metric does not consider outliers. Using the third quantile of localization errors, both estimates have an accuracy of 4.88 m.

These observations lead to the conclusion that several factors must be considered before deploying a CITS. Some of these factors are the number of participants willing to collaborate and the surface area of the deployment environment.

\subsection{Interest in using a similarity measure}
Unlike many papers presented in the literature, we chose the DFD as a metric for evaluating the performance of our proposed collaborative algorithm.
Highlighting trajectory similarities should be an essential performance metric for collaborative algorithms as it shows how location estimates diverge over time. This factor is essential as one of the most common use cases for indoor tracking is navigation. For this use case, the criteria for selecting a collaborative algorithm – and its parameters – must be how the location estimates diverge from the groundtruth.

\subsection{Limitations and future work}
The collaborative algorithm proposed in this paper is well-suited for rapid deployment in large areas as it does not require any infrastructure and uses collaboration to improve the accuracy of location estimates.
However, one of the limitations we do not cover in this paper is the human aspects of adopting such a solution. For instance, the reluctance of users to collaborate must be studied thoroughly.
A solution to address this limitation was proposed by Pourabdollah et al. They recommend offering extra services to reward users actively participating in the collaboration process or obliging non-collaborative users to pay a subscription fee to use the service~\cite{pourabdollah_lowcost_2010}.
As discussed in Section~\ref{subsec:performance_in_dense_environment}, we consider an environment where devices often encounter other devices and collaborate. In the future, we intend to consider environments where collaboration happens less often because there are fewer devices or few of them are willing to collaborate.

\subsection{Datasets}
The experiments and the evaluation can be reproduced using the data available on the following repository: \url{https://github.com/doplab/DCIT-Evaluation}.
\section{Related work}
\label{sec:related_work}

This section presents the related work addressing the challenges of building an ITS for infrastructure-free environments.
These solutions are either built with a decentralized architecture or use some collaboration.

Most of the solutions built for infrastructure-free environments are either motivated by afforbable-cost\cite{pourabdollah_lowcost_2010,pascacio_collaborative_2022}, ease of deployment~\cite{noh_infrastructure-free_2018} and the ability to deploy on most smartphones~\cite{ho_decentralized_2020}.
These three factors justify using widely available technologies such as wireless communication interfaces (BLE, WiFi, etc.) or IMU. Thus, these solutions target mobile devices with some processing power to run the tracking algorithms.
To tackle the performance issues that may arise from the restricted computing capabilities of these devices, some authors also introduced the notion of collaboration in their decentralized systems.

This is the case for Ho et al., who proposed a decentralized ITS based on BLE~\cite{ho_decentralized_2020}.
Their solution integrates two logical entities in the tracking process: anchor and target nodes. Anchor nodes, defined in our paper as trackers, are BLE beacons that can scan and broadcast information. Target nodes, defined as trackees, can be smartphones carried by a human, a robot or an IoT sensor. 
Their paper presents a way of encoding information into advertising data sent by trackers at predefined locations. Their solution eliminates the use of a central server, as the coordinates of the anchors are sent directly to the trackees via Bluetooth. In addition, the trackers exchange data collaboratively with each other. 
After receiving signals from nearby trackers, the trackers calculate the average signal strength from other anchors.
One of the limitations of their solution is the requirement of a set of beacons that cover a deployment environment. This approach is impractical in real life because BLE beacons are challenging to maintain, and using such devices significantly increases the cost of the solution. 
Another limitation is the requirement of BLE beacons with computing capabilities to broadcast, scan, store and analyze positioning data. The essence of BLE is power efficiency. That is the main reason why manufacturers design BLE beacons with the sole purpose of broadcasting data at predefined intervals.

In the literature, common use cases justifying the deployment of a decentralized CITS are time-critical team operations such as firefighting, urban military and search/rescue missions~\cite{noh_infrastructure-free_2018,morrison_collaborative_2017,liu_cooperative_2020}.
In such a constrained environment, installing any infrastructure is not feasible. That is why most decentralized CITS built for such scenarios use inertial sensors (i.e. dead reckoning) to compute the trackees' positions and exchange data in a peer-to-peer approach.

Noh et al. proposed a decentralized CITS for infrastructure-free environments~\cite{noh_infrastructure-free_2018}. Their solution works by generating an RSS map obtained with the path loss values of every coordinate on a map computed with a ray-tracing-based simulation. 
During the tracking process, each device obtains its position by performing WiFi beaconing through which the RSS values of the reachable devices can be observed. These values are then matched against the RSS map to generate a list of feasible coordinates. 
To correct the errors in positioning, each device uses dead-reckoning with inertial sensors to eliminate all the false positives from the list of feasible coordinates. If the obtained estimates lead to a dead-end, the invalid estimates are removed from the list, and the tracking process continues until each member gets unique coordinates. 

Like our proposed decentralized CITS, their solution offers the advantage of relying only on embedded sensors from the devices involved in the tracking process. However, their solution shows some limitations that need to be considered. The first limitation is that constructing an accurate RSS map with an RF ray-tracing model is challenging even in outdoor environments~\cite{vitucci_ray_2015}. Another limitation is the constant communication required for running the tracking algorithms. In their paper, the devices communicate via WiFi beaconing, which is unsuitable for most tracking systems as WiFi consumes much power and might deplete the devices' battery quickly~\cite{sadowski_rssi-based_2018}.

In the literature, collaboration for tracking is not only experimented on mobile devices. A growing community of researchers uses collaboration for Simultaneous Localization and Mapping (SLAM) with Autonomous Vehicles~\cite{leung2011distributed,paull_communication-constrained_2015}. The literature demonstrates that collaboration among the devices decreases the communication overheads generated by centralized systems while effectively achieving positioning in challenging environments\cite{kachurka_weco-slam_2022}.
However, most of these solutions use sensors not widely available on mobile devices such as the Lidar or laser pointers~\cite{waniek_cooperative_2015}.

 Some solutions proposed in the literature are related to ours in the sense that they use collaboration to reduce the accumulated errors of PDR~\cite{qiu_crisp_2018,li_indoor_2013}. However, these solutions require a central server to operate. This is mainly because of the high computational cost of running their tracking algorithms.

Another approach presented in the literature consists of using a Kalman Filter to correct the position estimates by fusing the data from different sources affected by noise or prone to errors.
This approach is mainly used for solutions using sensor fusion to correct the estimates of single devices. 
For instance, it is commonly used for drift removal or for filtering RSSI signals by fusing estimates from wireless and inertial-based systems~\cite{el-naggar_indoor_2019,you_hybrid_2021,sheikh_enhanced_2023}. 
In a more complex environment, this approach can suffer from high computational complexity when processing high dimensional data with frequent updates, substantially impacting real-time positioning, especially if it runs on constrained devices.

The solution presented in our paper was tested and deployed in an infra-structure-free environment. In such a constrained environment, the initial location estimates are made only using embedded inertial sensors. Collaboration using wireless signals is used occasionally to correct estimation errors. 
In addition, our solution can be deployed on devices with limited computing capacity and battery life, hence the interest in reconciling efficiency in estimation while using as few resources as possible.

\section{Conclusion}
\label{sec:conclusion}

In this paper, we proposed a novel collaborative algorithm to improve the location estimates of Pedestrian Dead Reckoning. 
This algorithm consists of low-complexity geometric operations that can run on mobile devices with limited battery life and constrained computing capabilities.
By relying on mobile devices for running the tracking algorithms, we remove the burden of deploying costly infrastructures and the risks associated with storing sensitive information on a remote server.
To demonstrate the effectiveness of our proposed collaborative algorithm, we used real-life data in the experiments.
We analyzed the movement of participants and selected parameters to maximize the accuracy of the proposed algorithm.
By collecting both inertial data and groundtruths, we obtained accurate measurements highlighting the advantages of collaboration in an infrastructure-free environment.
In the future, it would be interesting to study how the general public might perceive such systems, especially their willingness to collaborate. We will also evaluate the privacy risks, from a user perspective, of exchanging location data with nearby devices.

\bibliographystyle{unsrt}  
\bibliography{main.bib}

\end{document}